\documentclass[conference]{IEEEtran}

\usepackage{ifpdf}
\ifCLASSINFOpdf
  \usepackage[pdftex]{graphicx}
  \DeclareGraphicsExtensions{.pdf,.jpeg,.png}
\else
  \usepackage[dvips]{graphicx}
  \DeclareGraphicsExtensions{.eps}
\fi

\usepackage[utf8]{inputenc}
\usepackage[T1]{fontenc}
\usepackage[french, english]{babel}
\usepackage{cite}
\usepackage{caption}
\usepackage{amsmath,amsfonts,amssymb}
\usepackage{subcaption}
\usepackage{array}

\DeclareMathOperator*{\argmin}{arg\,min}

\begin{document}
%
%
\title{Coalition Formation Algorithm of Prosumers in a Smart Grid Environment}

%
%
\author{Nicolas Gensollen, Vincent Gauthier, Michel Marot and Monique Becker \\
\IEEEauthorblockA{CNRS UMR 5157 SAMOVAR, \\
Telecom SudParis/Institut Mines Telecom\\
Email: \{nicolas.gensollen, vincent.gauthier, michel.marot, monique.becker\}@telecom-sudparis.eu}}

\maketitle

%
%
\begin{abstract}
In a smart grid environment, we study coalition formation of prosumers that aim at entering the energy market. It is paramount for the grid operation that the energy producers are able to sustain the grid demand in terms of stability and minimum production requirement. We design an algorithm that seeks to form coalitions that will meet both of these requirements:  a minimum energy level for the coalitions and a steady production level which leads to finding uncorrelated sources of energy  to form a coalition. We propose an algorithm that uses graph tools such as correlation graphs or clique percolation to form coalitions that meet such complex constraints. We validate the algorithm against a random procedure and show that, it not only performs better in term of social welfare for the power grid, but also that it is more robust against unforeseen production variations due to changing weather conditions for instance. 
\end{abstract}

\IEEEpeerreviewmaketitle

%
%
\section{Introduction}
\label{sec:introduction}

One of the key ideas in the smart grid revolves around the introduction of communication means inside the power grids. This could enable complex improvements in the energy management and leads progressively to a greener energetic system \cite{Ramchurn} \cite{WuHamedHuangBook2011}. Distributed energy resources (DER) such as wind turbines or photovoltaic panels are not supposed to emerge only in remote farms, but also in residential areas. Together with electric vehicles, and demand side management programs \cite{Samadi2014}, they will constitute the building blocks which will help to turn the today pure energy consumers into true actors of the grid operation \cite{Ramchurn}. Such agents that both consume and produce energy are ready to make concessions (appliances delays, V2G) to ensure grid stability, and are commonly called prosumers \cite{6883384,Ramchurn}.

There seems to be a clear consensus on the benefit of having bidirectional communication flows between the prosumers and the grid, as demand side management relies on such an architecture \cite{WuHamedHuangBook2011}. Nevertheless as the number of active prosumers is expected to rise, it is safe to assume there is a need for more complex communication patterns (prosumers grouping together into so-called "coalitions") that will help to decrease the communication burden and satisfy the multiple requirements of the power grid management. Formation of coalitions inside a smart grid environment could be applied to various types of agents such as self sufficient micro-grids \cite{Pahwa}, sizable and adjustable virtual power plants (VPP) \cite{Braun, Ramchurn}, or fleet of electric vehicles that back up the grid in emergency situations (V2G) \cite{Ramchurn}. These are only a few use cases where the coalition of multiple agents can enhance the grid reliability and in the mean time reduce the communication burden.

In this paper, we consider the case where prosumers wanted to be clustered together into coalitions in order to be able to produce enough power which may be sold to the grid. If the coalition's production is sufficiently high and stable, the coalition can negociate on a day by day basis with the grid by announcing each day an expected production for the next day. We focus on how to statistically improve the production stability by carefully forming coalitions of prosumers that have greater probabilities of staying in acceptable range of production. By using prediction techniques, we form coalitions that meet the contract values ranges proposed by the grid, enabling it to schedule its production accordingly. Meeting contract values for the prosumers is especially relevant in power trade conditions, where energy is traded based on day-ahead predictions. In this context, participants should try to minimize their prediction errors in order to maintain a stable state for the grid. They may even occur some penalties if the productions deviate from the initial contract values. However, it remains pertinent for both side to maintain the production as stable as possible : for the prosumer, the stability (with renewable energy sources) will impact its net gain, and for the power grid, it will help to maintain its reliability.

In this paper, we seek to form coalitions that are able to announce "high contract values with high reliability". We thus aim at:
\begin{itemize}
\item Building a realistic prosumer model with renewable production sources based on weather data (see section \ref{subsec:ProsumerModel}).
\item Define a coalition formation model that enables the grid to set some requirements under which any group of prosumers will be allowed to sell its production (see section \ref{subsec:Coalition})
\item Define a utility function that will satisfy the grid requirements and maximize the stability of the prosumer coalition productions (see section \ref{subsec:UtilityFunc})
\item Define a coalition formation algorithm (see section \ref{sec:forming})
\end{itemize}

In section \ref{sec:model}, we consider that the agent's production depends on meteorological conditions (wind turbines and photovoltaic panels (PV) as generators), various energy mix and preferences, and their appliances (loads). We use meteorological data (see section \ref{subsec:ProsumerModel}) to account for seasonal and daily variations in the prosumer's energetic profiles as well as realistic geographical correlations between different agents. With these ingredients, we are able to run realistic simulations and record the different output profiles as time-series.

In order to select groups of prosumers that lower the volatility of the coalition's productions, we define an algorithm based on a popular approach for stocks market clustering \cite{Mantegna1999}. We use a modified clique percolation algorithm (see section \ref{sec:forming}) that will enable us to expand the cliques as needed in order to form proper coalitions that fulfill both the grid requirements and lower the production volatility. Finally, section \ref{sec:results} provides some results and a conclusion in section \ref{sec:conclusion}.

%
%

\section{Related Work}
\label{sec:related}

Traditionally, forming coalitions in a pool of agents can be done either in a centralized way where a single central unit is responsible for all the computations or in a distributed way where agents have only local knowledges and take actions accordingly. It is of common use to represent the situation and assess the stability of the solution by using game theoretic tools. Some papers \cite{Saad2009} \cite{Luan2014} focus then on finding an optimal coalition structure giving a pool of autonomous self interested agents using distributed merge/split algorithms.

Tackling the stability issues of renewable DER, the TradeWind project \cite{Europe} simulated the impact of wind power on electricity exchange and cross-border congestions by using a flow-based market model. The idea revolves around identifying key european interconnections (already existant or not) in order "to make optimal use of the european spacial de-correlation of wind power". It was indeed shown that geographical aggregation provides smoothening effects and that the amount of prediction errors for wind power in a geographical region diminishes as the region size increases, especially for short forecast horizons.

On a narrower scale, \cite{Kota2011} study the formation, in a game theoretic setting, of virtual power plants (VPP) composed of multiple self-interested DER. Two requirements for the formation of virtual power plants are considered : the reliability of supply and the minimization of entities the grid has to deal with. From this, \cite{Kota2011} builds a pricing mechanism that encourages VPP to report true estimates of their aggregated production and penalizes prediction errors. A redistribution scheme of the VPP to the DER is also constructed such that the payoff allocation lies in the core of the game, meaning that no DER has an incentive to leave the coalition.

In this paper we do not consider stability against player defection but focus on a statistics-oriented definition of coalition stability. We want to form coalitions based on utilities that depend on statistical properties of time-series values. The setting is thus similar with some financial studies on stock exchanges, where researchers tried to find relevant correlated clusters of stocks based on their daily prices variations. There exists well-known algorithms such as K-means or hierarchical clustering that are traditionally used for such purposes. Nevertheless, despite their popularity, they do not appear particularly well suited for meaningfully translating complex correlation relationships between time series while clustering them, especially when these series exhibit complicated underlying patterns. In \cite{Mantegna1999} the author introduced an approach where stocks time-series of their daily log returns, are organized in a graph such that stocks exhibiting similar price fluctuation patterns are close to each other. This closeness notion is formalized with a similarity measure based on Pearson correlation coefficient ($ d_{ij} = \dfrac{1}{2}\sqrt{2(1-\rho_{ij})} $ or $ d_{ij} = 1 - \rho_{ij}^{2} $ ) that enables to weight the edges of the graph. 

$\epsilon$-graphs, consists in filtering edges based on their weight, only keeping edges whose weights are less than $ \epsilon $. In \cite{Garas2008, Onnela2004}, the authors studied the topological properties (average clustering, connectivity, relative number of cliques) of the correlation graph against those of growing random graphs, depending on the threshold $ \epsilon $. However, there is no well defined method to select a right tradeoff $ \epsilon $ as a function of the network topology.

Presented in this way, the time-series clustering task seems very close to graph community detection. Communities in networks are indeed often seen as groups of nodes exhibiting high internal densities of links as well as a low density across communities, and several topology oriented techniques for finding communities are present in the litterature (\cite{Newman2013} \cite{Girvan2002} \cite{Newman2013_2}). For our purpose where decorrelation is the closeness notion, such algorithms tend to have some difficulties because only a few inclusions of very high correlations can strongly affect the stability of a coalition. However, local cliques, where all nodes are linked, provide uncorrelated groups of nodes. There exists greedy heuristics based on cliques in the litterature such as clique percolation \cite{Lancichinetti} which realizes local optimizations of a fitness function and results in overlaping community structures. Detection of overlapping communities is actually a very active field of research, especially in social networks where a person might belong to several communities.  


%
%

\section{Model}
\label{sec:model}
\subsection{Prosumer model}\label{subsec:ProsumerModel}

Our major concern is to characterize a prosumer behavior with realistic patterns of consumption, production as well as realistic geographical correlations between them. Only wind turbines and PV will be used as generators throughout this paper, and we denote by $ \mathcal{A} = \{ a_{1},...a_{N} \} $ the set of prosumers. To account for the production of a prosumer, we rely on French's weather data from 2006 to 2012 sampled every 3 hours \cite{Infoclimat} (similar data are available for the United States covering 2010\cite{NCDC}). As shown in the first part of figure 1, the first step consists in discretizing the studied zone around well chosen weather stations and gathering three kinds of traces (wind speed, cloudiness, and temperature) that we organize in a structure we will refer to as "climate vector" in the following. We consider that climate vectors are constant over their area, meaning that, if two agents are in the same area, they are exposed to the same climate vector. We use existing DER models enabling conversion between weather variables and output powers (see power curves in \cite{Kota2011} \cite{windturbinemodel}). \cite{Dans2007} provides also a convenient way of using cloudiness traces as realistic degradation factors in a \textit{"clear blue sky"} solar model, which enables us to rebuild realistic solar irradiance traces for stations that do not provide such information directly.

Consumption behaviors are characterized by a model we designed, which takes into account two major cycles : 

\begin{itemize}
\item \textbf{Daily cycles} : Consumption is low during night, and higher during the day with two picks in the morning and evening. Some noise is added so that prosumers have similar but not identical cycles.
\item \textbf{Seasonal cycles} : Consumption is higher in the winter because of heating and low in the summer (air conditioning is not considered). Temperature traces are used for modeling these cycles.
\end{itemize} 

\begin{center}
\begin{figure}
\includegraphics[scale=0.37]{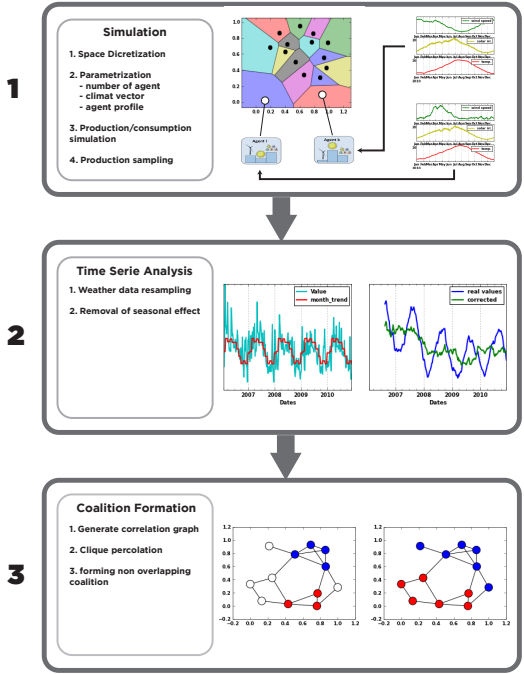}
\caption{Process diagram}
\label{Fig1}
\end{figure}
\end{center}

At this point, modeling agents consists in fixing a few parameters (most of the time drawn from random distributions) such as the geographical position, the number of wind turbines, PV, appliances, the temperature of confort, and so on. The objective is that, depending on the weather of his zone, the DER and appliances he owns, and the way he decides to heat his home, a given prosumer is able to compute his production/consumption at any time. 

Let $ P_{i}(t) $ be the instantaneous available power of agent i at instant t (his instantaneous production minus his instantaneous consumption). During the simulation (from $t_{0} $ to $ t_{K} $), all agents record their values with an hour time interval. As expected, the prosumer's time-series exhibit high seasonal patterns, but completely different from agent to agent because they depend both on the energetic mix and the habits. Nevertheless, these macro variations can hide the interesting information in the correlation coefficients and twist the rest of our process. Thereby, we remove these seasonal effects for all agents (see second block of figure \ref{Fig1}). We denote by $ \mathcal{T}_{i} = \{ P_{i}(t_{0}),...,P_{i}(t_{K}) \} $ the resulting time-series for  agent i, and by $ \mathcal{F}(P_{i}) $ the corresponding probability distribution.

\subsection{Coalition formation model}\label{subsec:Coalition}

Let us recall that our goal is to group prosumers into coalitions so that the global power production resulting of the superposition of individual prosumer's productions be sufficiently predictible and above a minimum power level. We thus extend the agent notations to any coalition $ s \subset \mathcal{A} $ : 
\begin{itemize}
\item $ P_{s}(t) = \sum_{i \in s} P_{i}(t) $
\item $ \mathcal{T}_{s} = \{ P_{s}(t_{0}),...,P_{s}(t_{K}) \} $ with $ \mathcal{F}(P_{s}) $ the corresponding probability distribution
\end{itemize}

Furthermore, we consider that a coalition S has the possibility to announce a contract value $ P_{s}^{CRCT} $ on the market. Due to the volatile nature of DER productions or user consumptions, the available power of a coalition will oscillate around $ P_{s}^{CRCT} $. It is of the responsibility of the information system that forms coalitions to select agents in order to provide a production as stable as possible. The probability of s value being below its contract value $ Pr[P_{s} \leq P_{s}^{CRCT} ] $ appears therefore as a quantity of interest that we would want to keep as low as possible. 

To illustrate what is done in the following, let us consider a very simple example with two agents, say i and j, with a production following a gaussian value probability distributions $ \mathcal{F}(P_{i}) = \mathcal{N}(\mu_{i}, \sigma_{i}) $ and $ \mathcal{F}(P_{j}) = \mathcal{N}(\mu_{j}, \sigma_{j}) $, such that the joint probability distribution  $ \mathcal{F}(P_{ij}) $ of the coalition $\{ij\}$ is also a gaussian with the following parameters :

\begin{equation}
\left\{ \begin{array}{lll}
		\mu_{ij} = \mu_{i} + \mu_{j} \\
		\sigma_{ij} = \sqrt{\sigma_{i}^{2} + \sigma_{j}^{2} + \rho_{ij}\sigma_{i}\sigma_{j}}
\end{array} \right.
\label{parameters}
\end{equation}

$ \rho_{ij} $ the Pearson correlation coefficient between $ P_{i} $ and $ P_{j} $. We can easily write the probability $ Pr[P_{ij} \leq P_{ij}^{CRCT} ] $ as :

\begin{equation}
\dfrac{1}{2} \left[ 1+ erf \left( \dfrac{P_{ij}^{CRCT} - \mu_{ij}}{\sigma_{ij}\sqrt{2}} \right) \right]
\label{reliability}
\end{equation}

Let $ \phi \in [0,1] $ be the \textbf{reliability} constraint imposed to the coalition $\{ij\} $. It stipulates that if $\{ij\} $ wants to join the market, the probability of $\{ij\} $ value (at any instant t) being below its contract value should at most be $ \phi $. That is, $ Pr[P_{ij} \leq P_{ij}^{CRCT} ] \leq \phi $. We also impose to $ \{ij\} $ a \textbf{minimum contract value} constraint denoted by $ P^{MIN} $, which states that the coalition should announce more than $ P^{MIN} $ to be accepted ($ P_{s}^{CRCT} \geq P^{MIN} $).

Coalition $ \{ij\} $ has thus potentially a set of possible contract values that respects both rules. The strategy of $ \{ij\} $ (and all other coalitions in the paper) consists then in choosing, if it exists, the maximum value of this constrained set. We denote by $ P_{\phi}(s) $ this maximum value for a coalition s. For $ \{ij\} $, $ P_{\phi}(s) $ can be computed using eq. \ref{reliability} : $ P_{\phi}(ij) = \mu_{ij} - \sigma_{ij}\sqrt{2}erf^{-1}(1-2 \phi ) $. If  $ P_{\phi}(ij) \geq P^{MIN} $, then $ \{ij\} $ can announce $ P_{ij}^{CRCT} = P_{\phi}(ij)$ on the market, else it is not allowed to enter. It thus appears (as it was intuitively understandable) that, for equivalent sizes, coalitions with low relative standard deviations ( $ \sigma_{ij} / \mu_{ij} $ ) are able to announce higher contract values. 

What this paper investigates in the following is the developement of a heuristic that organizes prosumers such that the synergy terms of the standard deviations ($ \rho_{ij}\sigma_{i}\sigma_{j} $ in the example above) are minimized. More formally, we consider the following problem (Table \ref{table1} summarizes the main notations):
 \begin{equation}
  \argmin _{\substack{ S \subset CS \\ 
                      |S| = N_{COAL} \\ 
                      \forall s \in S,\ |s| \neq 0 
                      \\ P_{s}^{CRCT} \geq P^{MIN}}}
   \sum_{s \in S}  \Pr \bigl[ P_{s} \leq P_{s}^{CRCT} \bigr]
  \label{problem}
\end{equation}  

As explained in the example above, if a coalition s wishes to join the market, it has to choose a contract value within a constrained set. Only coalitions whose set is not empty are able to enter the market (valid coalitions). For simplification, we consider that valid coalitions will always apply the same economically consistent strategy of announcing the highest possible contract value of this set ( $ P_{s}^{CRCT} = P_{\phi}(s) $).
Basically, a coalition s is valid if and only if :
\begin{equation}
\left\{ \begin{array}{lll}
		Pr[ P_{s} \leq P_{\phi}(s)] \leq \phi\ \textit{{\scriptsize (reliability rule)}} \\
		and\ P_{\phi}(s) \geq P^{MIN}\ \textit{{\scriptsize (min value rule)}}

\end{array} \right. 
\label{rules}
\end{equation}

\begin{table}[t]
\centering
\scriptsize
\begin{tabular}{l|p{4.5cm}}
  \hline 
  $ P_{i}(t) \in \mathbb{R}\ \ (P_{s}(t)) $ & Instantaneous production of agent $i$ at time $t$ \\
  $ \mathcal{F}(P_{i})\ \ (\mathcal{F}(P_{S})) $ & Probability distribution of $i$'s production \\
  $ \phi \in [0,1] $ & Reliability threshold \\
  $ P^{MIN} \in \mathbb{R}^+ $ & Min value threshold \\
  $ P_{s}^{CRCT} \in \mathbb{R}^+ $ & Contract value announced by coalition s \\
  $ P_{\phi}(s) \in \mathbb{R}^+ $ & Max contract value under $ \phi $ constraint \\
  $ \mathcal{U}_{\phi,\ P^{MIN}}(s) $ & Utility function \\
  $ \tau_{i}(s) \in \mathbb{R}$ & Overlapping to disjoint coalitions mapping \\
  $ \epsilon \in [0,1] $ & Pruning threshold for the correlation graph \\
  $ N_{COAL} \in \mathbb{Z}^+ $ & Number of desired coalition \\
  $ CS $ & Set of all coalition structures \\
  $ \Theta(G) $ & Set of non overlaping cliques in graph G \\
  \hline
\end{tabular}

\caption{Notations\label{table1}}
\end{table}

\subsection{Utility function}\label{subsec:UtilityFunc}
Adressing the problem of eq.\ref{problem}, we choose a utility function (eq. \ref{utility}) that derives directly from the above remarks. If a coalition cannot provide a valid contract value, it receives naturally a utility of zero. Furthermore, it seems obvious that the utility should increase with the contract value. The $ 1/|s| $ term in eq. \ref{utility} indicates that we favor small coalitions, mainly because they are easier to maintain in terms of communications.

\begin{equation}
 \mathcal{U}_{\phi,\ P^{MIN}}(s) = \mathbf{1}_{\textit{s\ valid}} \dfrac{P_{\phi}(s)}{|s|} 
\label{utility}
\end{equation}

Obviously, maximising this utility function amounts to maximizing the coalition contract value with the minimum possible number of agents. In such settings, and through the clique percolation procedure (section \ref{sec:forming}), we will show that coalitions with high utilities can be computed.

%
%

\section{Coalition Formation}
\label{sec:forming}

This section explains the process with which we form the coalitions (see the third block of Figure \ref{Fig1}). First, we need to simulate the time-series of available power (first two blocks of Figure \ref{Fig1}). We consider a pool $ \mathcal{A} $ of 200 agents, whose parameters were chosen randomly. The prosumers are positioned (also randomly) on a square lattice previously filled with climate vectors obtained from the french data sets (see section \ref{sec:model}). Simulations were run from february 2006 to december 2010 such that we are dealing with 200 hourly sampled time-series of available power over this period. Removing season trends finally leads us to the formation of coalitions. 

The model we used in order to simulate time-series of available production provides some diversity because of the combination between the energetic mix (renewable generators combination) and climate vectors. Nevertheless, as the number of agents grows, the time-series tend to exhibit similar patterns. This is apparent when creating a correlation graph with similar metric such as defined in \cite{Garas2008} or \cite{Onnela2004} ($ 1 - \rho_{ij}^{2} $), where well defined clusters appear in the $ \epsilon $-graph for any values of $ \epsilon $. 

However, these clusters of strongly correlated time-series are the exact opposite of what we are seeking. We can indeed consider them directly as coalitions and compute their utilities, and the results show (see the green curves on figure \ref{Fig4}), as expected, terrible values (far worse than a random split of the agents in the same number of coalitions). We thus opt for reversing the metric ($ \rho_{ij}^{2} $) such that uncorrelated time-series are close to each other in the graph and correlated time-series are distant. We defined a graph $G_\epsilon(\mathcal{A},E)$ where $\mathcal{A}$ is the set of agents and $ E $ the edge set. A given link between $ i $ and $ j $ is present if $ \rho_{ij}^{2} \leq \epsilon $, that is, $ e_{i,j} = \mathbf{1}_{\{\rho_{ij}^{2} \leq \epsilon \}}\rho_{ij}^{2} $. As expected \cite{Onnela2004}, independently of the $ \epsilon $ parameter, $ G_{\epsilon } $ exhibits henceforth much less clustering and communities seem hardly visible. Therefore, using classical clustering or community detection algorithms seem to provide poor results. 

However, as seems intuitively understandable, cliques of this graph tend to exhibit very good utility values. Such  structures contain indeed a link between every two nodes, meaning that the overall correlation is quite small. Obviously, the $ \epsilon $ parameter is indirectly responsible for the sizes of the cliques : if it is too low, $ G_{\epsilon} $ does not provide enough cliques, conversely, if $\epsilon $ is too high, we loose important information as the graph becomes very dense and cliques tend to overlap strongly. For large values of $ \epsilon $ and for non trivial number of agents, finding cliques can even become computationally intensive. Despite being direct and simple, improving the sizes of the cliques by increasing $ \epsilon $ seems too brutal. Furthermore, as the utility function is also focused on maximizing $ P_{\phi}(s) $, it might be the case that some cliques benefit from additional agents even if they don't form a clique anymore. Cliques for small values of $ \epsilon $ appear thus as good seeds for stable coalitions. We define $ N_{COAL} $ as the desired number of coalitions fixed by the user. Furthermore, let $ \Theta(G_{\epsilon}) $ be the set of non overlaping cliques in $ G_{\epsilon} $, we choose the $ \epsilon $ parameter as the smallest value such that $ G_{\epsilon} $ contains at least $ N_{COAL} $ non overlaping cliques :

\begin{equation}
\epsilon^{\star} = \min_{\substack{ \epsilon \in [0,1] }} \Big\{ \epsilon\,\ s.t\ |\Theta(G_{\epsilon})| \geq N_{COAL} \Big\}
\end{equation}

We construct $ G_{\epsilon^{\star}} $ such that it contains at least $ N_{COAL} $ potential seeds that may or may not be above the grid requirements, but exhibit low correlations among the members. Naturally, grid requirements could eventually be reached and social welfare increased by incorporating more nodes into the seeds. For this purpose, we use a clique percolation algorithm on $ G_{\epsilon^{\star}} $ that will extend the seeds over a correlation constrained environment. The main idea behind this algorithm is to make the seeds grow by local optimization of a fitness function, usually based on internal and external degrees of the seeds \cite{Lancichinetti}. The main difference between our case study and this algorithm for community detection is that we chose the fitness function as our utility function (cf. eq.\eqref{utility}) that both maximizes the production and minimizes its volatility. As explained in section \ref{sec:related}, clique percolation leads generally to overlapping communities. For simplicity, in this paper, we wish to keep the coalitions disjoint and  leave the management of overlapping coalitions for future work. We thus implemented a simple heuristic that considers nodes in multiple seeds one by one and chooses its final coalition as the one that “needs it the most” in terms of utility loss. More formally, for a coalition s and a node $ i \in s $, we define :

\begin{equation}
\tau_{i}(s) = \dfrac{\mathcal{U}_{\phi,\ P^{MIN}}(s) - \mathcal{U}_{\phi,\ P^{MIN}}(s-\{i\}) }{\mathcal{U}_{\phi,\ P^{MIN}}(s)}
\label{tau}
\end{equation}

If node i belongs to  multiple coalitions, the only coalition including node i is the one that maximizes $ \tau_{i} $. At this point, we have three degrees of freedom : the reliability ($ \phi $), the required power to enter the market ($P^{MIN}$), and the number of desired coalitions ($ N_{COAL} $). Next section shows the utility behavior within the parameters space $ \{\phi,P^{MIN},N_{COAL} \} $ and provides results of our 200 agents test.

\section{Results}
\label{sec:results}

\begin{figure}
 \centering
  \includegraphics[scale=0.7]{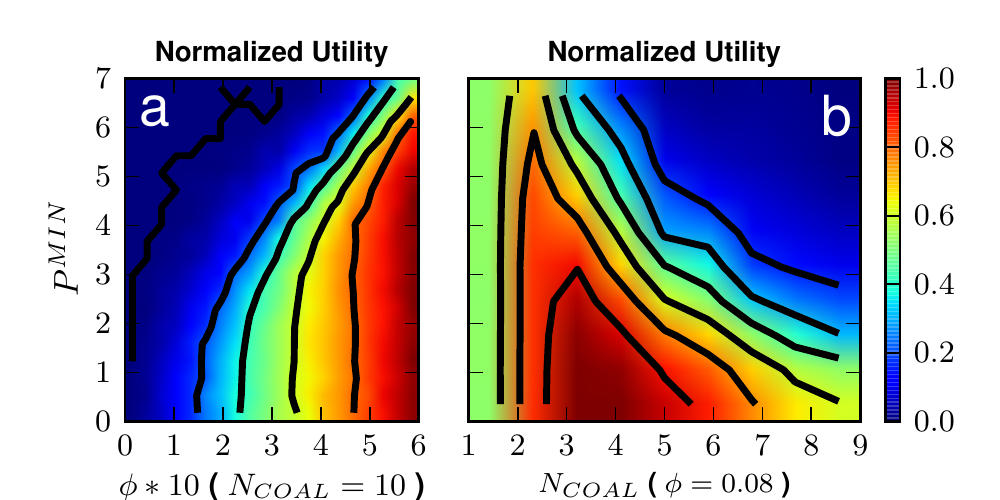}
\caption{Utility as function of the parameters space $\{\phi \in [0,1], N_{COAL}\in \mathbb{Z}^+, P^{MIN} \in \mathbb{R}^+\}$: a) In $ \{\phi, P^{MIN}\} $ when $ N_{COAL} $ is fixed. b) in $ \{N_{COAL}, P^{MIN}\} $ for a fixed $ \phi $. For readability, $ P^{MIN} $ is expressed in tenth of MW.}
\label{Fig2}
\end{figure}

As shown in figure \ref{Fig2}a, the parameters $\phi$ and $P^{MIN}$ shape the utility function such that, if $ \phi $ is close to zero, the reliability requirement is very high. Only small values of $ P^{MIN}$ could then potentially lead to valid coalitions (and positive utilities). Conversely, the higher $\phi$, the less constraints are imposed to the coalitions and more valid coalitions can arise for a larger spectrum of $ P^{MIN}$. Obviously, the highest utility values are found for high $ \phi $, because the formed coalitions are able to announce higher contract values, yielding higher utilities. 

In figure \ref{Fig2}b, we fix the reliability to a given empirical value ($\phi = 0.08 $). We can observe how the social welfare $\sum_{s \in S} \mathcal{U}_{\phi,\ P^{MIN}}(s)$ evolves according to $P^{MIN}$ and $ N_{COAL} $. In our setting we notice a maximum point around $ N_{COAL} = 3 $, followed by a decrease. There is clearly a maximum number of coalitions sustainable for a given set of parameters. After that value the coalitions newly formed won't be able to pass the grid requirements. Moreover, reckon that increasing $ N_{COAL} $ means also increasing $ \epsilon $, leading to denser graphs, meaning that the algorithm performance will also decrease.

\begin{figure}[htbp]
  \centering
  \includegraphics[scale=0.6]{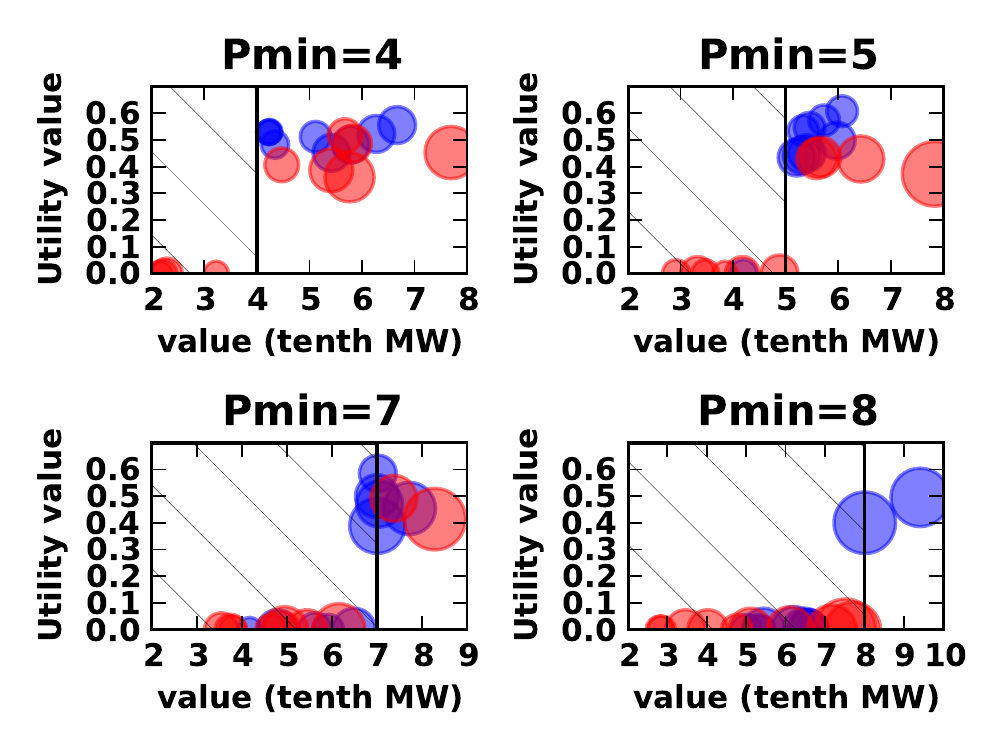}
  \caption{Evolution of the coalitions for different values of $ P^{MIN} $ for clique percolation (blue dots) and random process (red dots). The diameters of the dots are proportional to the coalition sizes. The hatched zero utility zone corresponds to the \textit{"under-requirement space"}.}
  \label{Fig3}
\end{figure}

In fig. \ref{Fig3}, we perform a simulation with 200 prosumers, and we fix the reliability $ \phi = 0.1 $ meaning that coalitions should produce more than their contract values at least $ 90 \% $ of the time. We observe the result of our algorithm for a fixed number of coalitions of $N_{COAL} = 10$ and let the algorithm form the coalitions as the grid changes according to $ P^{MIN} $. We compare our algorithm with a random process that splits the agents in $ N_{COAL} $ coalitions. Fig. \ref{Fig3} shows this evolution for our algorithm (blue dots) and for the random process (red dots). The diameter of a dot is proportional to the number of agents in the coalition and the higher the dot, the higher its utility. The $ P^{MIN} $ values of the x axis are expressed in tenth of MW for readability and the hatched zone corresponds to the \textit{"under-requirement space"}, meaning that whenever a coalition is in this zone, it has a null utility. Looking at fig. \ref{Fig3}, we can see first that our algorithm performs better than the random formation algorithm at finding high valued coalitions. The blue dots allowed to enter the market (non hatched zone) are indeed outnumbering the red dots, especially when the grid requirements are neither too low nor too high ($ P^{MIN} = 5 $ in fig. \ref{Fig3}).

\begin{figure}
 \centering
  \includegraphics[scale=0.3]{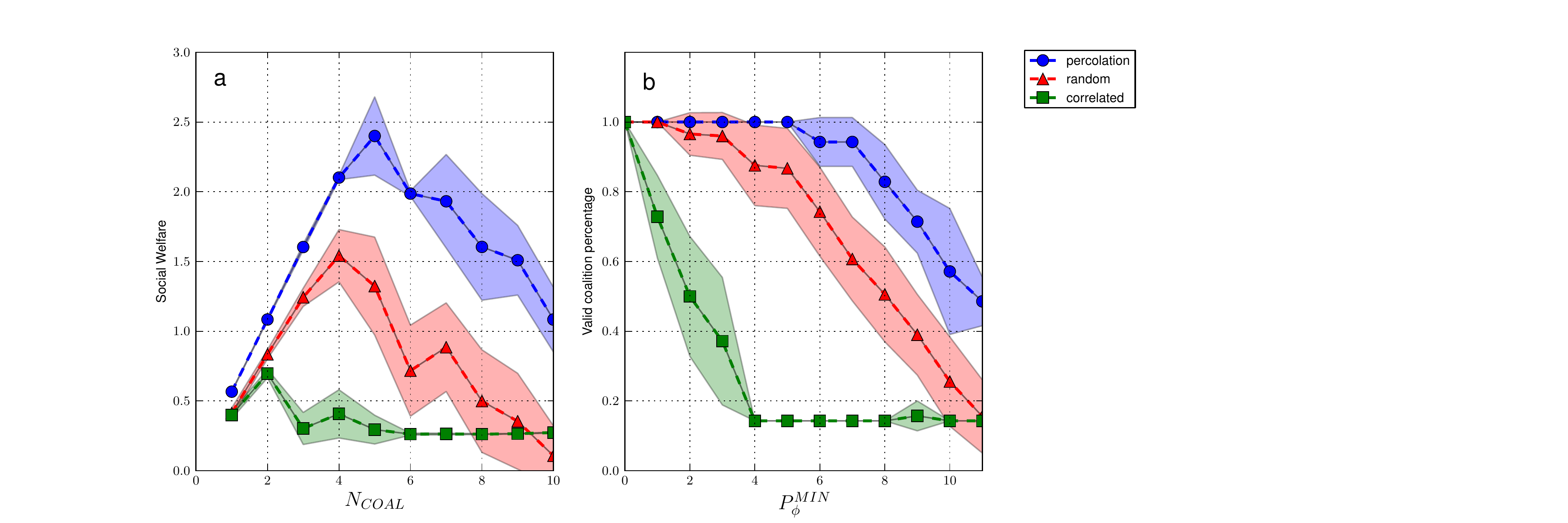}
  \caption{Evolution of the social welfare \textbf{a)} and the percentage of coalitions able to enter the market \textbf{b)} for different values of $ P_{MIN} $. Blue curves represent clique percolation, red curves, the random process, and the green ones, correlated coalitions.}
 \label{Fig4}
\end{figure}

In more details, figure \ref{Fig4}a presents the evolution of social welfare as the number of coalitions increases (all other parameters are kept constant) for random (red curve), clique percolation (blue curve), and correlated coalitions (green curves) that stands for a worst case scenario. As for figure \ref{Fig4}b, it shows the percentage of coalitions able to enter the market for different values of $ P^{MIN} $. For consistency, we average the results of both plots over 100 realizations (colored zones around the curves shows the standard deviations). When the grid requirements are constant (figure \ref{Fig4}b), and the number of desired coalitions is low, clique percolation generally performs better than random case. Moreover, when $ N_{COAL} $ gets bigger, the performance of a random split tumble down rapidly while our clique percolation is able to maintain efficiently the social welfare of the coalitions formed. When the grid requirements vary, for very low $ P^{MIN} $, all coalitions for all algorithms are able to enter the market, yielding an acceptance percentage of $ 100 \% $. But as $ P^{MIN} $ increases, we see the percentage of the correlated coalitions is going down very quickly. After a few increases in $ P^{MIN} $, the percentage of the random procedure starts dropping while it stays almost constant for our algorithm. For $ P^{MIN} = 8 $, we see that only a little more than half of the coalitions for the random case are able to enter while approximately $ 85 \%$ of them enters for the clique percolation algorithm. Finally, when the grid requirements becomes too high, the acceptance percentage of our algorithm tends slowly to zero (not shown in this plot for readability).

%
%
\section{Conclusion}
\label{sec:conclusion}

We believe that the originality of this paper lies in its willingness to exploit the de-correlation of prosumer profiles in order to build stable coalitions. In this direction, we presented a model based on meteorological traces, that captures the complex "\textit{energetic mix/climate vectors}" combination and generates realistic production and consumption patterns. We then built a framework that enables the grid to specify stability and minimum production requirements for filtering the coalitions. On this basis, we proposed a simple algorithm that seeks for uncorrelated prosumer patterns as potential seeds and expand them in coalitions able to rise above the grid requirements. We validated our algorithm against a random choice of coalitions, which can be considered as an average case. We showed  that it performs better (coalitions are more stable and the overall production is more important) and that it exhibits a higher robustness/flexibility against grid requirements changes. The worst case scenario is represented by coalitions of correlated prosumers. Interesting leads for future works would be the use of correlated clusters to reduce the number of entities the algorithm has to deal with, or the introduction of a payoff allocation towards the prosumers such that the stability against player defection could be analyzed through game theory. Showing that maintaining the coalitions formed with our algorithm necessitates less communication and less storage capacity could also conduct to a stimulating project. Besides, we believe that not restricting the algorithm to non overlapping coalitions and studying the strategies and weights of nodes with multiple options could lead to interesting works.

%
%
 
\bibliographystyle{IEEEtran}  
\bibliography{Article}

\begin{thebibliography}{10}
\providecommand{\url}[1]{#1}
\csname url@samestyle\endcsname
\providecommand{\newblock}{\relax}
\providecommand{\bibinfo}[2]{#2}
\providecommand{\BIBentrySTDinterwordspacing}{\spaceskip=0pt\relax}
\providecommand{\BIBentryALTinterwordstretchfactor}{4}
\providecommand{\BIBentryALTinterwordspacing}{\spaceskip=\fontdimen2\font plus
\BIBentryALTinterwordstretchfactor\fontdimen3\font minus
  \fontdimen4\font\relax}
\providecommand{\BIBforeignlanguage}[2]{{%
\expandafter\ifx\csname l@#1\endcsname\relax
\typeout{** WARNING: IEEEtran.bst: No hyphenation pattern has been}%
\typeout{** loaded for the language `#1'. Using the pattern for}%
\typeout{** the default language instead.}%
\else
\language=\csname l@#1\endcsname
\fi
#2}}
\providecommand{\BIBdecl}{\relax}
\BIBdecl

\bibitem{Ramchurn}
S.~D. Ramchurn \emph{et~al.}, ``Putting the 'smarts' into the smart grid: A
  grand challenge for artificial intelligence,'' \emph{Commun. ACM}, vol.~55,
  no.~4, pp. 86--97, Apr. 2012.

\bibitem{WuHamedHuangBook2011}
C.~Wu \emph{et~al.}, \emph{Smart Grid Communications and Networking},
  E.~Hossain \emph{et~al.}, Eds.\hskip 1em plus 0.5em minus 0.4em\relax
  Cambridge University Press, 2012.

\bibitem{Samadi2014}
P.~Samadi \emph{et~al.}, ``{Utilizing renewable energy resources by adopting
  DSM techniques and storage facilities},'' \emph{International Conference on
  Communications}, pp. 4221--4226, 2014.

\bibitem{6883384}
L.~Xiao \emph{et~al.}, ``Anti-cheating prosumer energy exchange based on
  indirect reciprocity,'' in \emph{International Conference on Communications},
  2014, pp. 599--604.

\bibitem{Pahwa}
S.~Pahwa \emph{et~al.}, ``{Topological Analysis and Mitigation Strategies for
  Cascading Failures in Power Grid Networks},'' 2010.

\bibitem{Braun}
M.~Braun, ``{Virtual Power Plants in Real Applications Pilot Demonstrations in
  Spain and England as part of the European project FENIX},'' 2009.

\bibitem{Mantegna1999}
R.~Mantegna, ``\BIBforeignlanguage{English}{Hierarchical structure in financial
  markets},'' \emph{\BIBforeignlanguage{English}{The European Physical Journal
  B}}, vol.~11, no.~1, pp. 193--197, 1999.

\bibitem{Saad2009}
W.~Saad \emph{et~al.}, ``A distributed coalition formation framework for fair
  user cooperation in wireless networks,'' \emph{Wireless Communications, IEEE
  Transactions on}, vol.~8, no.~9, pp. 4580--4593, September 2009.

\bibitem{Luan2014}
X.~Luan \emph{et~al.}, ``Cooperative power consumption in the smart grid based
  on coalition formation game,'' in \emph{Advanced Communication Technology
  (ICACT)}, Feb 2014, pp. 640--644.

\bibitem{Europe}
F.~Van~Hulle \emph{et~al.}, ``{Integrating Wind, European project report},''
  2009.

\bibitem{Kota2011}
G.~Chalkiadakis \emph{et~al.}, ``Cooperatives of distributed energy resources
  for efficient virtual power plants,'' in \emph{The 10th International
  Conference on Autonomous Agents and Multiagent Systems}, 2011, pp. 787--794.

\bibitem{Garas2008}
A.~Garas \emph{et~al.}, ``\BIBforeignlanguage{English}{The structural role of
  weak and strong links in a financial market network},''
  \emph{\BIBforeignlanguage{English}{The European Physical Journal B}},
  vol.~63, no.~2, pp. 265--271, 2008.

\bibitem{Onnela2004}
J.-P. Onnela \emph{et~al.}, ``\BIBforeignlanguage{English}{Clustering and
  information in correlation based financial networks},''
  \emph{\BIBforeignlanguage{English}{The European Physical Journal B -
  Condensed Matter and Complex Systems}}, vol.~38, no.~2, pp. 353--362, 2004.

\bibitem{Newman2013}
M.~E.~J. Newman, ``{Spectral methods for network community detection and graph
  partitioning},'' p.~11, Jul. 2013.

\bibitem{Girvan2002}
M.~Girvan \emph{et~al.}, ``Community structure in social and biological
  networks,'' \emph{Proceedings of the National Academy of Sciences}, vol.~99,
  no.~12, pp. 7821--7826, 2002.

\bibitem{Newman2013_2}
M.~E.~J. Newman, ``{Community detection and graph partitioning},'' no.~2, p.~5,
  May 2013.

\bibitem{Lancichinetti}
A.~Lancichinetti \emph{et~al.}, ``Detecting the overlapping and hierarchical
  community structure in complex networks,'' \emph{New Journal of Physics},
  vol.~11, no.~3, p. 033015, 2009.

\bibitem{Infoclimat}
\BIBentryALTinterwordspacing
 [Online]. Available: \url{http://www.infoclimat.fr}
\BIBentrySTDinterwordspacing

\bibitem{NCDC}
\BIBentryALTinterwordspacing
 [Online]. Available: \url{http://www.ncdc.noaa.gov/}
\BIBentrySTDinterwordspacing

\bibitem{windturbinemodel}
\BIBentryALTinterwordspacing
 [Online]. Available: \url{http://www.wind-power-program.com}
\BIBentrySTDinterwordspacing

\bibitem{Dans2007}
G.~Dans \emph{et~al.}, ``{Rayonnement solaire},'' vol. 360, 2007.

\end{thebibliography}
\end{document}